\documentclass[aps,twocolumn,floats,floatfix,showpacs,superscriptaddress]{revtex4}
\usepackage{epsfig}
\begin{document}
\title{On the Widespread Occurrence of the Inverse Square Distribution in 
Social Sciences and Taxonomy}
\author{Guido Caldarelli}
\affiliation{INFM UdR ROMA1 Dipartimento di Fisica, Universit\`a di Roma
``La Sapienza'' Piazzale A. Moro 2 00185 Roma, Italy}
\affiliation{Istituto di Studi e Ricerche "E. Fermi" Compendio Viminale 00184 Roma, Italy}
\author{C\'ecile Caretta Cartozo}
\affiliation{INFM UdR ROMA1 Dipartimento di Fisica, Universit\`a di Roma
``La Sapienza'' Piazzale A. Moro 2 00185 Roma, Italy}
\author{Paolo De Los Rios}  
\affiliation{Laboratoire de Biophysique Statistique, ITP-FSB, Ecole Polytechnique F\'ed\'erale
de Lausanne, CH-1015 Lausanne, Switzerland}
\author{Vito D.P. Servedio} 
\affiliation{INFM UdR ROMA1 Dipartimento di Fisica, Universit\`a di Roma
``La Sapienza'' Piazzale A. Moro 2 00185 Roma, Italy}
\date{\today} 
\begin{abstract} 
The widespread occurrence of an inverse square relation in the hierarchical distribution 
of sub-communities within communities (or sub-species within species) has been 
recently invoked as a signature of hierarchical self-organization within social 
and ecological systems. Here we show that, whether such systems are self-organized or not, 
this behavior is the consequence of the tree-like classification method. 
Different tree-like classifications (both of real and truly random systems) 
display a similar statistical behaviour when considering the sizes of 
their sub-branches.
\end{abstract} 
\maketitle

\begin{figure} 
\centerline{\psfig{file=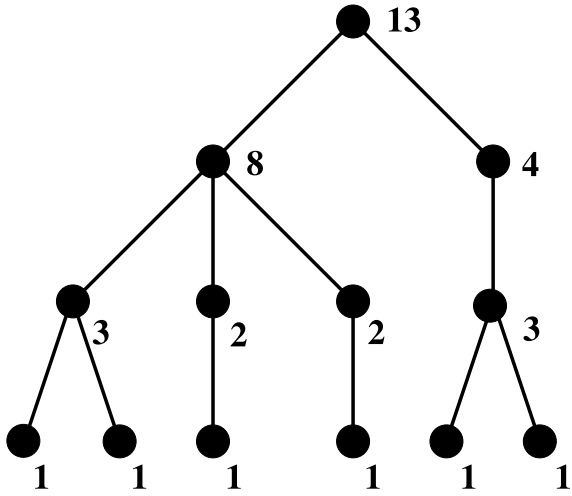,width=7.5cm}} 
\caption{Binary tree with nodes labelled by their sub-tree sizes.}
\label{Fig1}
\end{figure}

Taxonomy is one of the major classification schemes in use in all natural and social sciences.
The main underlying assumption driving a taxonomic class partition is that the entities under study 
are hierarchically organized: for example, in biology four different {\it species}
({\it e.g.} lions, tigers, lynxes and cats) can be grouped into two {\it genera} (lions and
tigers into Pantherinae, lynxes and cats into Felinae), which in turn belong to the same 
{\it family} (Felidae); in social sciences different sub-communities can be grouped in 
the same community and so on: supporters of different soccer teams can be grouped, at a 
higher level, as soccer fans that in turn, together with fans of other disciplines, 
can be grouped as sports fans and so on.
Therefore, taxonomic trees carry information about the laws of organization of 
organisms and communities.  In addition, in the last decade
much work has been devoted to the study of the statistical features of taxonomic trees,
in the hope that they could reveal some general 
patterns of organization.  Interestingly, when looking at the statistical distribution 
of the sizes of sub-trees (which correspond to the sizes of subspecies within species, 
of subcommunities within communities) an inverse square power-law relation
has been found\cite{Bur,Bur2,Loreto}. 
Subtree sizes are best viewed by Fig.\ref{Fig1}, where we show a tree and 
label every node by the size of its corresponding subtree. 
It is worth noticing that such scheme 
corresponds to the statistics of drainage basins area in the river 
network studies\cite{rinaldo}.
The presence of a power-law in the size distribution of sub-tree
sizes has been hailed as a consequence of some self-organization
(possibly critical) of the 
system under consideration. 
Actually, and even more interestingly, the robustness of such a law across different 
taxa and kingdoms (in ecology) and for different social contexts and 
definitions of
community (in social sciences) points to some universal property of 
these systems.

Here we show that, although self-organization is surely present in ecological 
and social systems, its presence is not revealed by the above mentioned inverse square relation,
which instead is an 
unescapable consequence of the tree-like nature of classification scheme. 
Moreover, the robustness of the inverse square relation emerges simply as a 
universal consequence of the tree-building algorithm.

Community detection within a society is not a simple and perfectly defined task. 
Indeed, the definition itself of a community is not clear-cut, and even for a 
given definition 
the procedure to identify communities might not be easy to implement. 
Quite recently Girvan and Newman (GN) have proposed an algorithm that, on test cases
(where the communities are already known) seems to perform quite effectively
\cite{GN}. 
The algorithm relies on a simple intuition: communities are groups of individuals tightly 
connected with each other, and less connected
to the rest of the social network. Therefore, by identifying a suitable parameter 
that quantifies such degree of connectedness, it is possible to devise a procedure 
to fragment the social network in communities. 
GN use the edge betweenness~\cite{freeman} index to estimate how much an edge 
of the network is important to keep it connected: in general edges within a 
community have a low betweenness value, since the global connectivity
is anyway ensured by the large amount of edges linking individuals in the 
community. Edges between two different communities, on the other hand, 
should exhibit high betweenness values, since there should
be a few of them, sustaining all the social relations between the 
two communities. The GN algorithm recursively prunes the network 
by cutting the edge with the largest betweenness, checking whether the network has split in 
two fragments, and repeating the procedure untill the network
is made of disconnected nodes. Any time there is a splitting, the taxonomic 
tree of communities is updated: the community that has split into two subcommunities 
gives rise to two offsprings. The algorithm therefore
builds a binary tree that encodes the community structure of the social network.
More recently, other algorithms\cite{newmann,Loreto} have been proposed that run faster 
than the GN one by relying on local quantities, such as the edge clustering (the computation
of the betweenness is costly since it is a global quantity). Yet, the common feature is that
the algorithm still produces binary taxonomic trees.
As mentioned above, all these algorithms correctly identify communities in selected test cases, and all
exhibit an inverse square law behavior in the community size distribution, sometimes
used as evidence of a self-organized hierarchical structure in the system. Yet, the structure,
encoded in a taxonomic tree, is hierarchical by construction, and also the inverse square power behavior
is an inescapable outcome of the construction. 

\begin{figure}
\centerline{\psfig{file=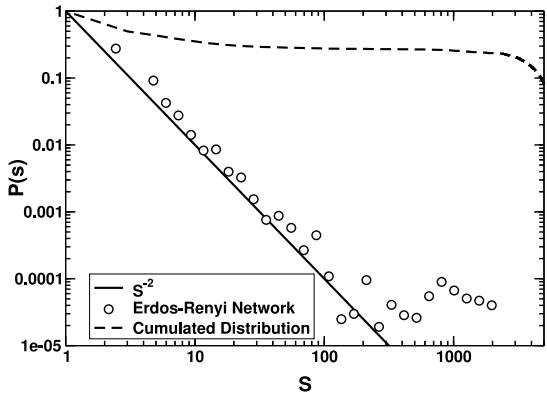,width=9.0cm}}
\caption{Community size distribution for an Erdos-Renyi network of $3000$ nodes and
with average degree equal to $10$. The straight line is
$s^{-2}$.}
\label{Fig2}
\end{figure}

It has been analytically shown recently that the
sub-tree size distribution (the community size distribution in the 
social context) of a random tree is always power-law distributed, 
$P(s)\sim s^{-\gamma}$, with the exponent $\gamma$ that can be equal
to $3/2$ in the case of critical random trees~\cite{harris} (a critical random tree 
is a tree where nodes have, on the average, just a single offspring), 
and $2$ in the more common case of supercritical
trees (the average number of offsprings is larger than one)\cite{delos}.
To check whether this result applies in general to the detection algorithm, 
we have applied the one based on the edge clustering coefficient to a simple 
Erdos-Renyi network made of $3000$ nodes and with
an average degree equal to $10$. As shown in Fig.\ref{Fig2}, 
the community size distribution shows an inverse power-law behavior for almost 
two decades. Yet, the network is completely random, and there is no 
self-organization whatsoever in the communities identified by the algorithm.
The application of the betweenness based GN algorithm gives similar results (data not shown). 
In Fig.\ref{Fig2} we have logarithmically binned the data: this is helpful to improve the quality
of the plot, although not essential to reveal the power-law in the community size distribution (the non binned data
are just slightly noisier). 
The cumulated distribution, instead, wouldn't have shown any of these features, as Fig.\ref{Fig2}
clearly shows: this is a {\it caveat} in the use of cumulated distributions, that can sometimes
heavily distort the results, to the point that by their use
the universality of the inverse square power relation could have been missed.

\begin{figure}
\centerline{\psfig{file=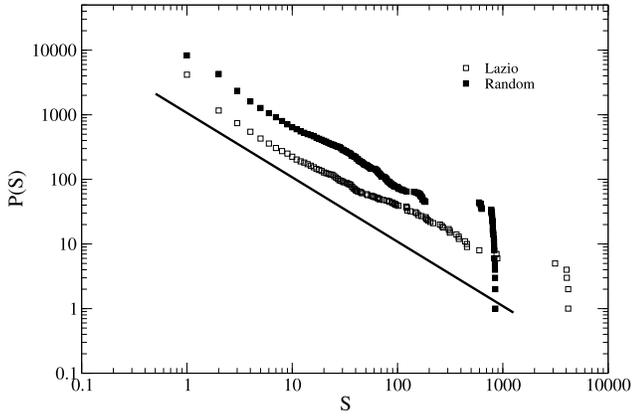,width=7.0cm,angle=270}}
\caption{Community size distribution for the flora of Lazio and a
random subset of italian species. The straight line is $s^{-1}$ since these
are cumulated distributions.}
\label{Fig3}
\end{figure}

The inverse square-law behavior also emerges from real data and from sets of real
data after some randomization. As an example we examine data from
plant taxonomy, which is
a particular case of biological classification.
We have considered a number of plants ecosystems
belonging to different climatic and geographical environments
around the world, each one represented by the taxonomic classification
tree. For all these trees the study of the statistical distribution
of the sizes of subtrees shows a very good inverse power-law distribution
with an exponent $\gamma\sim 1.9\pm 0.2$. As all the plants species in
each tree belong to a same ecosystem, this result could be viewed as a
sign of some self-organization of the ecosystem itself.  
Yet, in this case a simple experiment shows 
that community statistics do not reveal any structure (although the details of
the community partition do). 
We have studied two different kinds of subsets of the above 
ecosystems: geographical subsets made of  groups of species
that live toghether in a restricted area~\cite{flora} and give rise to a small ecosystem, 
and some other completely random subsets. 
The statistical distribution of the
sizes of subtrees shows an inverse square power-law behavior
for both kinds of subsets (see Fig.\ref{Fig3}). 
Finding this behavior both in homogeneous data from the same
ecosystem, and in a random subsets of species where no correlations are present,
shows once more that the inverse square-law is simply a consequence
of the tree-like nature of the taxonomic classification.
It is worth observing, in this case, that Fig.\ref{Fig3} shows the
cumulated distribution, that in this case preserves its power-law features
(this is further evidence that, if the cumulated distribution is a
power-law then the original is power-law too, whereas the inverse is
not necessarily true, for example if data are quite noisy).
 
In conclusion we have shown that the presence of an inverse square-law
relation for the distribution of the sizes of communities and species
does not mark any significative organization within the 
system under consideration, but rather it is an inevitable
consequence of the classification scheme. 
More generally we point out that such power-law behavior should be 
expected whenever the classification scheme
used to organize a set of data is based on the construction of a 
hierarchical tree. This should
also apply to classical taxonomy, where the appearance of power-laws 
in the distributions of
subspecies within species has been linked to some intrinsic 
fractality in the way species are organized.
The degree of correlation and organization
of the system under study should be instead tied to the quality
of the scaling rather than to its bare presence. These
subtle relations are yet to be discovered.
In some cases, the community size distribution has shown power-law
scaling with exponent $3/2$, rather then $2$\cite{Guimera}: since it is known
that the only ingredient necessary to find such a scaling for random trees
is their criticality (number of offsprings equal to $1$), we speculate that
it is not the presence of scaling itself that points to some peculiarities
in these systems, but rather the specific value of the exponent (and we stress
that no other exponents different from $2$ and $3/2$ could be found).

We thank the EU FET Open project IST-2001-33555 COSIN, 
the OFES-Bern under contract 
02.0234.


\end{document}